V. V. Obukhovsky[*], Doct. Sci., prof.,
V. V. Nikonova[**], post grad. stud.

# Nonlinear diffusion in Acetone-Benzene Solution

*The nonlinear diffusion in multicomponent liquids under chemical reactions influence has been studied. The theory is applied to the analysis of mass transfer in a solution of acetone-benzene. It has been shown, that the creation of molecular complexes should be taken into account for the explanation of the experimental data on concentration dependence of diffusion coefficients. The matrix of mutual diffusivities has been found and effective parameters of the system have been computed.*

[*]E-mail: vobukhovsky@yandex.ru
[**]E-mail: vika_p_06@mail.ru

## Introduction

The diffusion in liquids is usually described by standard diffusion equation (Fick's law) [1]. Numerous experimental data indicates that in «binary» solutions the Fick's diffusion coefficients $D_F$ are not constant, but depend on solute concentration [1-2]. Up to now there is not general theory to explain in details the dependence of the diffusivities versus solvent concentration for many-components liquids.

It is known the chemical reactions accompany processes of dissolving. In particular, due to intermolecular interactions [4, 5] "molecular complexes" can occur. But effect of these processes on diffusivity is investigated only a little.

The macroscopic version of the nonlinear diffusion theory was developed in [5-7] for liquid solutions that take into account the reaction between the mixture components. The purpose of this work is theoretical studying of diffusion in manycomponent solutions. The example of acetone-benzene solution is analyzed in details.

## 1. Diffusion in liquid solution

Let's consider the diffusion in molecular solutions, that were formed by mixing of two substances *X* and *Y*. Interaction between molecules of the original components leads to the formation (with some probability) complexes of the type *[$X^n Y^m$]*. In further confine our investigations by systems with reactions

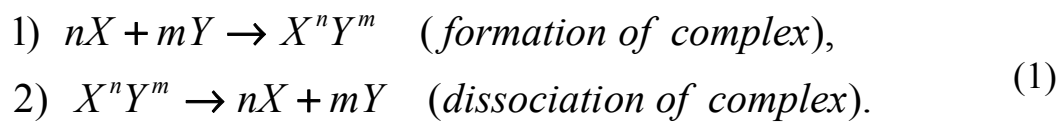

$$\begin{aligned}&1)\ nX + mY \rightarrow X^n Y^m \quad (formation\ of\ complex),\\&2)\ X^n Y^m \rightarrow nX + mY \quad (dissociation\ of\ complex).\end{aligned} \quad (1)$$

(To simplify the calculations, the intermediate steps of reactions are not considered).

In the frames of our model, this kind of mixture consists of three components: $X$, $Y$ and $[X^n Y^m]$. Hereinafter these components are denoted as 1, 2, 3. The basis of the nonlinear diffusion theory [5-7] is conception: the diffusion transfer of macroscopic amounts of matter in one direction is accompanied by a countercurrent of other substances inevitably (it is necessary for conservation of the total volume of mixture).

Since individual molecules can vary in size widely, it makes sense to formulate a theory in terms of "partial volumes" $M_n$. Under this approach, the diffusion is described by next laws [5-7]:

- equations of continuity

$$\frac{\partial M_i}{\partial t} + div\, \vec{j}_i = S_i, \qquad (2)$$

- nonlinear diffusion flows

$$\vec{j}_i = \sum_j d_{ij}[M_j \nabla M_i - M_i \nabla M_j]. \qquad (3)$$

Here the following notation were used: the indices $i, j$ numerate the components of the liquid mixture; $M_i$ – volume fraction of components $i$ ("partial volume"); $\vec{J}_i$ – volume's flow of $i$-substance; $S_i$ – function of sources, which depend on the processes of $i$-component birth (decay); $d_{ij}$ - mutual diffusion coefficients, that forms matrix $\{d_{ij}\}$.

It can be shown that the nonlinear form of the flow (3) in particular case does not contradict the linear Fick's law (see Appendix A). Equations (2-3) are valid in the range $0 \leq M_n \leq 1, (n = 1, 2, ...)$. In the case of negligible small influence of shrinkage/swelling phenomenon on the processes of diffusive transport, we take into account the law of volume conservation:

$$\sum_{i=1}^{n} M_i = 1. \qquad (4)$$

Reactions (1) determine the form of functions $S_i$ (sources):

$$S_1 = \beta_1 M_3 - \alpha_1 M_1^n M_2^m,$$
$$S_2 = \beta_2 M_3 - \alpha_2 M_1^n M_2^m, \qquad (5)$$
$$S_3 = -S_1 - S_2.$$

Below it is supposed the chemical reactions proceed quickly, but the spatial mass transfer – slowly enough. It corresponds to the condition

$$S_i \cong 0. \qquad (6)$$

Another words, diffusion occurs when local chemical equilibrium has been took place. In this case it follows from (5):

$$M_3 \cong \gamma M_1^n (1-M_1-M_3)^m, \quad \left(\gamma = \frac{\alpha 1}{\beta 1} = \frac{\alpha 2}{\beta 2}.\right) \tag{7}$$

Under diffusion process molecules are transported both individually and as part of the complex $[X^n Y^m]$. Therefore, the total flow of matter "$X$" (measurable in experiments) is defined as a linear combination:

$$\vec{j}_1^{total} = \vec{j}_1 + \eta_1 \vec{j}_3 \tag{8}$$

(Here $\eta_1 = n\Delta V_1/(n\Delta V_1 + m\Delta V_2)$ - volume fraction of substance $X$ in complex $[X^n Y^m]$). Excluding $M_2$ we can find

$$\vec{j}_1^{total} = [-d_{12} + q_0 M_3]\nabla M_1 + [-\eta_1 d_{23} - q_0 M_1]\nabla M_3, \tag{9}$$

where $q_0 = (d_{12} - d_{13}) + \eta_1(d_{13} - d_{23})$.

As we pointed out above, the substance 1 can be in solution in two states: free and bound (part of molecular complex composition). Therefore, its total volume is

$$M_1^{total} = M_1 + \eta_1 M_3, \tag{10-а}$$

Similarly, we can determine $M_2^{total}$:

$$M_2^{total} = M_2 + \eta_2 M_3, \left(\eta_2 = \frac{m\Delta V_2}{n\Delta V_1 + m\Delta V_2}\right). \tag{10-б}$$

In this case the normalization condition automatically execute

$$M_1^{total} + M_2^{total} = 1. \tag{10-в}$$

It is easy to verify assertion: the value $M_1^{total}$ completely determines all other functions, if conditions (6) take place: i.e. $M_i = M_i(M_1^{total})$, $(m=1,2,3)$. Therefore the total flux of substance 1 also can be represented as a function of $M_1^{total}$ only:

$$\vec{j}_1^{total} = -D^{ef}(M_1^{total})\nabla M_1^{total}. \tag{11}$$

It is obvious, equation (11) has the form corresponds to the law of Fick's diffusion with generalized (effective) coefficient of diffusion $D^{ef}$. The last is depending on the "concentration" of interacting substances. If the reactions (sources) are described by (5), one can found

$$D_1^{ef}(M_1^{total}) = \frac{(d_{12} - q_0 M_3) + \Phi(\eta_1 d_{23} + q_0 M_1)}{1 + \eta_1 \Phi}, \tag{12-а}$$

where

$$\Phi = \frac{M_3[n(1-M_1-M_3) - mM_1]}{M_1(1-M_1-M_3 + mM_3)}. \tag{12-б}$$

The results (12) describe the diffusion in the liquid mixture with components $X, Y, [X^n Y^m]$. Material parameters of this system are the following: $\gamma, d_{nm}, \Delta V_n$.

**2. Analysis of diffusion in acetone - benzene mixture**

Let's apply a modified system of diffusion equations (2-5) to analyze the mass transfer processes in a liquid mixture of acetone with benzene. These substances are mutually soluble completely. Experimental data on diffusion in acetone-benzene mixture were obtained at $T = 25°C$ [8-9].

In further acetone *(A)* is considered as component 1, and benzene *(B)* as a component 2. Enthalpy of mixing $\Delta H$ for these substances is nonzero (see Fig.1 based on data from [3]).

The $\Delta H$ peak corresponds to a molar ratio 1:1 (or one molecule of acetone, on average, to one molecule of benzene). It can be interpreted as the result of reaction $A + B \rightarrow C$, where $C \equiv [A^1 B^1]$. Hence, in this case *n=1, m=1*. Molecules of acetone or benzene do not change chemically in reaction of this kind, but "attached" to each other (with some probability). Thus, the initially binary system (before mixing) in the process of reaction/diffusion may be considered as a ternary (triple).

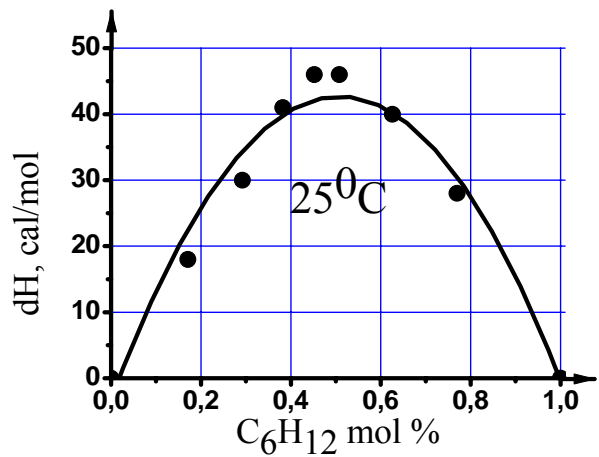

Fig. 1 Enthalpy of acetone-benzene mixing at $25^0C$ [3]

Under chemical reactions, one portion of the molecules $A, B$ remain in a free state, but another is spend on formation of molecules $C$. Partial volumes of the substances (that proportional to concentrations) satisfy the condition (4).

The standard flow (in particles/cm$^2$s) as function of usual concentrations (in particles/cm$^3$) was investigated in [7-8]. These experimental data have been restated to construct a graph of the diffusion coefficient $D^{\exp}$ as a function of the acetone partial volume $(M_1^{total})$. Theoretical curve $D_1^{ef}(M_1^{total})$ for acetone-benzene mixture was calculated according to (12) numerically. The following material parameters

were found from the condition of the best agreement between theoretical curve and experimental results:

$$d_{12} = 2.18, \quad d_{13} = 7.24, \quad d_{23} = 3.83, \quad \gamma = 1.17 \quad . \tag{13}$$

Volume of molecules

$$\Delta V_1 = 122.6 \left( \overset{0}{A} \right)^3, \quad \Delta V_2 = 148.6 \left( \overset{0}{A} \right)^3$$

were calculated according to data [3]. Theoretical results with experimental data are presented in Fig.2.

The good agreement theory-experiment can be interpreted as a confirmation of the assumption: the coefficients $d_{nm}$ are constants and do not depend on concentrations.

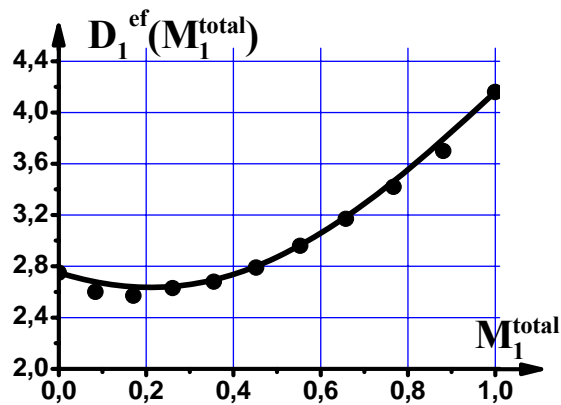

Fig. 2 Concentration dependence of the effective diffusion coefficient. Points - experimental data [10-11], solid curve - theoretical calculation.

Parameters (13), which were found from diffusion measurements, permit to solve the additional task: to find the concentration of molecular complexes. The results of numerical simulation are presented in Fig.3.

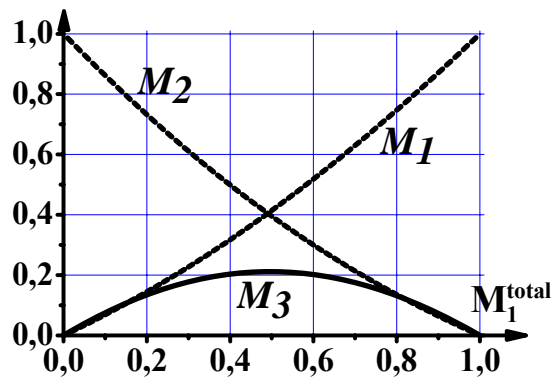

Fig. 3 Concentration dependence of partial volumes acetone-benzene solution.

Values of $M_i$ are proportional to concentrations of components.

**Conclusions**

In result we can conclud:

- three-component model $\{A, B, C\} = \{X, Y, [X^n Y^m]\}$ can be successfully applied to modeling of diffusion in mixtures, where only one type of complexes is formed,

- non-monotonic dependence of the diffusion coefficient versus the initial concentration of acetone can be explained by generation of the molecular complex in acetone-benzene mixture. Its structure is similar to $[A^1 B^1]$.

It should be noted, the generalization of this theory on shrinkage/swelling phenomenon accounting can be done similarly to [10].

## Appendix A

Let's consider a mixture of two non-interactive molecular liquids. From the condition of volume conservation $M_1 + M_2 = 1$ follows:

$$M_2 = 1 - M_1, \quad \nabla M_2 = -\nabla M_1, \tag{A-1}$$

Substituting (A-1) to (3) we find the expression for the "volume flow"

$$\vec{j}_1 = -d_{12}\nabla M_1. \tag{A-2}$$

Diffusion coefficient in (A-2) is constant. If each molecule of substance 1 occupies a volume $\Delta V_1$, from (A-2) one can get the diffusion equation in the standard form of Fick's law

$$\vec{J}_1 = -D_F^0 \nabla N_1, \tag{A-3}$$

Here $\vec{J}_1 = \vec{j}_1 / \Delta V_1$ - the number of particles (molecules) of type 1, which cross the unit surface area per unit time, and $N_1 = M_1 / \Delta V_1$ – the concentration of these particles.

## Appendix B

In extreme cases ($M_1^{total} = 0$, $M_1^{total} = 1$,) effective diffusion coefficient (12) and its first derivative can be written as:

$$D_1^{ef}\left(M_1^{total} = 0\right) = \frac{d_{12} + (\gamma \cdot \eta_1)d_{23}}{1 + (\gamma \cdot \eta_1)}, \tag{B-1}$$

$$D_1^{ef}\left(M_1^{total} = 1\right) = \frac{d_{12} + (\gamma \cdot \eta_2)d_{13}}{1 + (\gamma \cdot \eta_2)}, \tag{B-2}$$

$$\frac{dD_1^{ef}}{dM_1^{total}}\left(M_1^{toalt} = 0\right) = +2(\gamma\eta_1)(1+\gamma)\frac{(d_{12} - d_{23})}{(1 + (\gamma \cdot \eta_1))^3}, \tag{B-3}$$

$$\frac{dD_1^{ef}}{dM_1^{total}}\left(M_1^{total} = 1\right) = -2(\gamma\eta_2)(1+\gamma)\frac{(d_{12} - d_{13})}{(1 + (\gamma \cdot \eta_2))^3}. \tag{B-4}$$

The left part on (B-1)-(B-4) can be determined from the experimental data. Four parameters $d_{12}$, $d_{13}$, $d_{23}$, $\gamma$ can be found by solving the system of four equations by numerical methods.

# References

1. Diffusion in Condensed Matter. Ed. by P.Heitjans, J.Kärger. 2005, pp.966, Springer-Verlag.
2. *John J. Mc. Ketta*. Encyclopedia of Chemical Prosessing and Design – N.-Y. – 1985.
3. *Белоусов В.П., Морачевский А.Г.* Теплоты смешения жидкостей. Л.: Химия ЛО. – 1970. – 256 с.
4. *Пиментел Джордж, К. Мак-Клеллан, Обри Л.* Водородная связь. – М.: Мир – 1964. – 462 с.
5. *E.S. Gyulnazarov, V.V. Obukhovsky, T.N. Smirnova*. Theory of holographic recording on a photopolymerized material //SPIE "Milestone Seria". – 1996. – **130**. – P. 473-475.
6. *G.M.Karpov, V.V.Obukhovsky, T.N.Smirnova, V.V.Lemeshko*. Spatial transfer of matter as a method of holographic recording in photoformers // Opt. Communication. – 2000. –**174**, № 5-6. – P. 391-404.
7. *Ю.В. Гріднева, В.В. Обуховський*. Вплив міжмолекулярної взаємодії на дифузію в рідинах // Вісник Київського університету. – 2003. – №3. – С. 284-288.
8. *Chaohong He*. Prediction of the calculation dependence of mutual diffusion coefficients in binary liquid mixtures. //Eng. Chem. Res. 1995 **34**. – P.2148-2153.
9. *David W. McCall, Dean C. Douglass*. Diffusion in binary solutions. //J. Phys. Chem. 1967. - **71**. – P. 987-997.
10. *H.M. Karpov, V.V.Obukhovsky, T.N. Smirnova*. Generalized model of holographic recording in photopolymer materials //Semiconductor Physics, Quantum Electronics & Optoelectronics. – 1999. –**2**, №3. – P. 66-70.